\documentclass[apj,twocolumn]{openjournal}

\usepackage{lipsum}
\usepackage{newtxtext,newtxmath,enumitem,multirow}

\usepackage{xcolor}
\usepackage{textgreek}
\usepackage[utf8]{inputenc}
\usepackage[english]{babel}
\usepackage{graphicx}	
\usepackage{amsmath}	
\usepackage{amssymb}	
\usepackage{wasysym}    

\usepackage[T1]{fontenc}

\usepackage{booktabs}
\usepackage{placeins}
\usepackage{capt-of}
\usepackage{csvsimple}
\usepackage{orcidlink}
\usepackage{xspace}

\usepackage{hyperref}
\hypersetup{
    colorlinks=true,
    linkcolor=blue,
    citecolor=blue,
    filecolor=magenta,      
    urlcolor=cyan,
    pdfpagemode=FullScreen,
    }

\makeatletter 
  \patchcmd{\NAT@citex}
    {\@citea\NAT@hyper@{%
      \NAT@nmfmt{\NAT@nm}%
      \hyper@natlinkbreak{\NAT@aysep\NAT@spacechar}{\@citeb\@extra@b@citeb}%
      \NAT@date}}
    {\@citea\NAT@nmfmt{\NAT@nm}%
    \NAT@aysep\NAT@spacechar\NAT@hyper@{\NAT@date}}{}{}

  \patchcmd{\NAT@citex}
    {\@citea\NAT@hyper@{%
      \NAT@nmfmt{\NAT@nm}%
      \hyper@natlinkbreak{\NAT@spacechar\NAT@@open\if*#1*\else#1\NAT@spacechar\fi}%
        {\@citeb\@extra@b@citeb}%
      \NAT@date}}
    {\@citea\NAT@nmfmt{\NAT@nm}%
    \NAT@spacechar\NAT@@open\if*#1*\else#1\NAT@spacechar\fi\NAT@hyper@{\NAT@date}}
    {}{}
\makeatother

\newcommand{\uvot}{{\it Swift}-UVOT}
\newcommand{\xrt}{{\it Swift}-XRT}
\newcommand{\swift}{{\it Swift}}

\shorttitle{SAPLE}
\shortauthors{Marcotulli \& Torres-Alb\`a}

\begin{document}

\title{SAPLE: Swift Analysis Pipeline for Lightcurve Extraction\vspace{-1.3cm}}

\author{Lea Marcotulli$^{1,2}$$\orcidlink{0000-0002-8472-3649}$,
N\'uria Torres-Alb\`a$^{2,3,\dagger}$$\orcidlink{0000-0003-3638-8943}$
}

\affiliation{
$^{1}$Deutsches Elektronen-Synchrotron DESY, Platanenallee 6, 15738 Zeuthen, Germany\\
$^{2}$Department of Physics and Astronomy, Clemson University, Kinard Lab of Physics, Clemson, SC 29634-0978, USA \\
$^{3}$Department of Astronomy, University of Virginia, Charlottesville, 22904, VA, USA \\
$^{\dagger}$GECO Fellow
}

\begin{abstract}

We present the \swift~Analysis Pipeline for Lightcurve Extraction (SAPLE), a semi-automated pipeline to extract the \uvot~and \xrt~data products and spectral information (magnitudes, photon indices, and fluxes) for a set of observations of any {\it point source} of interest. This pipeline is not meant to substitute, but to complement the tools the Swift team has already set up.  Specifically, SAPLE provides a \uvot~semi-automated pipeline that also returns the absorption corrected specific fluxes for any observation and filter of interest, a tool which to our knowledge is not publicly available to the community yet. Moreover, for \xrt, SAPLE enables the user to extract a lightcurve of both flux and photon index (with associated uncertainties), assuming a redshifted powerlaw spectrum. The main codes are available through a GitHub repository \citep{marcotulli2026}, and the following paper summarizes the main steps of the analysis.

\end{abstract}
\maketitle


\section{Introduction} 

In orbit since 2004, the Neil Gehrels \swift~Observatory has been (and still is) an extremely successful mission. Onboard, it hosts 3 instruments: the ultra-violet and optical monitor (UVOT, 170-650 nm, \citealp{Roming_2005_UVOT}), the X-Ray Telescope (XRT, 0.3-10 keV, \citealp{Burrows_2005_XRT}), and the Burst Alert Telescope (BAT, 14-195 keV, \citealp{Barthelmy_2005_BAT}). Built originally as a multi-wavelength mission to catch and study gamma-ray bursts (GRBs), and monitor the GRB afterglow evolution, the scientific reach of the mission has surpassed its primary goal (with $>120$ papers published per year since its launch\footnote{\url{https://www.swift.ac.uk/results/pub.php}}). One of \swift's strongest characteristics is its design to slew and repoint very fast to any target of opportunity (now down to the order of $\sim10$ seconds via an API system, \citealp{Tohuvavohu_2024}), and, as such, \swift~has been dominating the time domain UV and X-ray astronomy field for the last 2 decades.  

With time-domain and multi-messenger astrophysics being at the fore-front of today's scientific landscape, missions like \swift~are of paramount importance for high-energy astrophysics. \xrt~was indeed the first instrument to {\it repoint}  to the first gravitational wave event with a clear electromagnetic counterpart, GW170817 \citep[within $\sim1$ hour from the gravitational wave signal alert,][]{Evans_2017}, and to the neutrino alert of TXS 0506+056 \citep[in $\sim3.25$ hours after the neutrino alert,][]{txs_2018} and identify the multi-wavelength counterparts of such events. Moreover, with the collection of more than 22 years of data,  about 14\% of the entire sky has been observed by XRT and UVOT, with 95\% of sources reported in the Living \xrt~Point Source (LSXPS) catalogue observed more than once \citep[$N_{\rm obs}>2$,][]{Evans_2023}, giving the community a long term baseline of optical-to-X-ray monitoring of sources of interest. The \swift~team has also developed tools publicly available to the community to analyze data, specifically: the \textit{Swift-XRT data products generator}\footnote{\url{https://www.swift.ac.uk/user\_objects/docs.php}}, which enables the creation of X-ray images, spectra, count-rate and hardness ratio light curves, and the position calculation of any point source in the \xrt~ field of view; the \texttt{BatAnalysis} Pipeline \citep{BATPipeline_2025}, which allows the analysis of data from the BAT Survey and Time-tagged Event (TTE) to extract light curves and spectra for each observation for a given source, as well as to create mosaiced images at different time bins; and, importantly, an extensive and constantly updated documentation for data analysis for all three instruments\footnote{\url{https://www.swift.ac.uk/analysis/index.php}}.  

With the goal of analyzing a large number of observations from blazar sources \citep{Penil_2024a, Penil_2024b, Penil_2026}, we built a set of tools to semi-automatically analyze the \xrt~and \uvot~data of the targets and extract spectral information. The major driver behind developing these ad-hoc tools was that, as of today, (i) we did not find (to the best of our knowledge) an automated UVOT tool that reduces the \uvot~data and provides spectral information of point sources, and (ii) that the \xrt~tools do not provide the photon index or flux light curve information, but only the count rate/hardness ration based ones (plus allows for different types of time binning, a function which we did not implement). 

Therefore, here we release SAPLE: \textit{Swift Analysis Pipeline for Lightcurve Extraction} available through the Github repository, \citet{marcotulli2026}. This set of codes provides the user with:

\begin{enumerate}
    \item A \uvot~semi-automated pipeline that returns the magnitudes and \textit{absorption corrected} specific fluxes for all desired observations of the {\it point source} of interest;
    \item A \xrt~semi-automated pipeline that extracts lightcurves of both \textit{absorbed} and \textit{absorption corrected} flux and photon index (with associated uncertainties), assuming a redshifted powerlaw spectrum, of the {\it point source} of interest.
\end{enumerate}

In Section~\ref{sec:prerequisite} we outline the prerequisite needed to start working with SAPLE; in Section~\ref{sec:data-analysis} we describe the standard data processing that SAPLE relies on; in Section~\ref{sec:artifcats} we enumerate the artifact and issues one may encounter while working with \uvot~ and \xrt~ pointings that need to be manually dealt with in SAPLE; in Section~\ref{sec:fluxes} we detail how the fluxes are extracted for both instruments; in Section~\ref{sec:products} we present the deliverable of SAPLE and one astrophysical examples and in Section~\ref{sec:future} we list some of the future developments for SAPLE.

\section{Prerequisites}\label{sec:prerequisite}
The main prerequisite to using SAPLE is the fact that the user has to have a working version of HEASoft\footnote{\url{https://heasarc.gsfc.nasa.gov/docs/software/lheasoft}} (\negthickspace\citealp{heasoft_2014}) and the most up-to-date CaLibration DataBase (CALDB\footnote{\url{https://heasarc.gsfc.nasa.gov/docs/heasarc/caldb}}) files for \xrt~and \uvot~installed in the machine they will be using for the analysis. 
Importantly, the XRT fitting codes use the pyXSPEC \citep{Gordon_2021} implementation of the X-ray spectral (Xspec, \citealp{xspec_1996}) fitting tool, hence it is up to the user to ensure pyXSPEC works before even starting the data analysis. The python packages needed for the codes to run are listed in the GitHub folder. 

\section{Data Analysis Steps}\label{sec:data-analysis}

For all the observations you will be dealing with, SAPLE follows the standard \xrt~and \uvot~steps detailed in the \swift~guide to data reduction\footnote{\url{https://www.swift.ac.uk/analysis/index.php}}, which we describe in the next sub-sections.  

\subsection{UVOT}
The workflow of the UVOT data analysis pipeline is outlined as follows:
\begin{enumerate}
    \item First, the user creates a source and background region for one of the UVOT observation. The particular choice of observation is not important, as long as the target is visible. The same source and background regions -- i.e., with identical shape, size and coordinates -- are used as starting point for all analyzed observations. However, for each exposure, the code automatically loads the base source region, centroids it on the position of maximum counts of the target, and saves it as a new source region specific to that observation.  
    The suggested shape/size of the source region is a circle of $5''$. For the background, it is a circle of $\sim15''$, which should be located close to the source but away from other contaminants. Note that these are just recommended shapes/sizes, and the user is free to chose what fits best for their analysis or observation. 
    \item Using DS9, the pipeline produces PNG images for all observations and filters, enabling visual inspection to confirm appropriate source and background region selection and the absence of artifacts in each exposure. The user is in charge to manually remove all the exposures that may present issues (see Section~\ref{sec:artifcats});
    \item Finally, the \texttt{uvotsource} task is run on all observations and filters to extract the optical-UV magnitudes of the target. 
\end{enumerate}

\subsection{XRT-PC and XRT-WT}
The XRT data analysis is run on both Windowed Timining (WT) and Photon Counting (PC) events and is structured as follows: 
\begin{enumerate}
    \item At first, the \texttt{xrtpipeline} is run on all the observations downloaded by the user;
    \item Then, the user creates source and background regions for one of the XRT exposures. 
    The same source and background regions -- i.e., with identical shape, size, and coordinates -- are used as the starting point for all analyzed observations. However, for each exposure, the code automatically loads the base source region, centroids it on the position of maximum counts of the target, and saves it as a new source region specific to that observation.  
    For \textit{PC events}, the user can select the shape and size of both regions as they see fit (the team has tested SAPLE with circular source and background regions). For \textit{WT events}, both the source and background region shapes are fixed as circles with the same fixed radius ($r$, as per XRT-WT guidelines): the source region is centered at the source coordinates; the background is centered $2\times r$ from the source region position along the slit. The user can only change the size of the radius (the default one in SAPLE is $r=50''$); 
    \item Using DS9, the pipeline produces PNG images for all observations and filters, enabling visual inspection to confirm appropriate source and background region selection and the absence of artifacts in each exposure. The user is in charge of manually removing all the exposures that may present issues (see Section~\ref{sec:artifcats});
    \item Finally, the \texttt{xselect} task is run on all observations to extract the unbinned spectra for the source and the background.
\end{enumerate}

\section{Artifacts \& Issues}\label{sec:artifcats}
\subsection{UVOT}

\begin{figure*}
    \centering
    \includegraphics[width=0.6\linewidth]{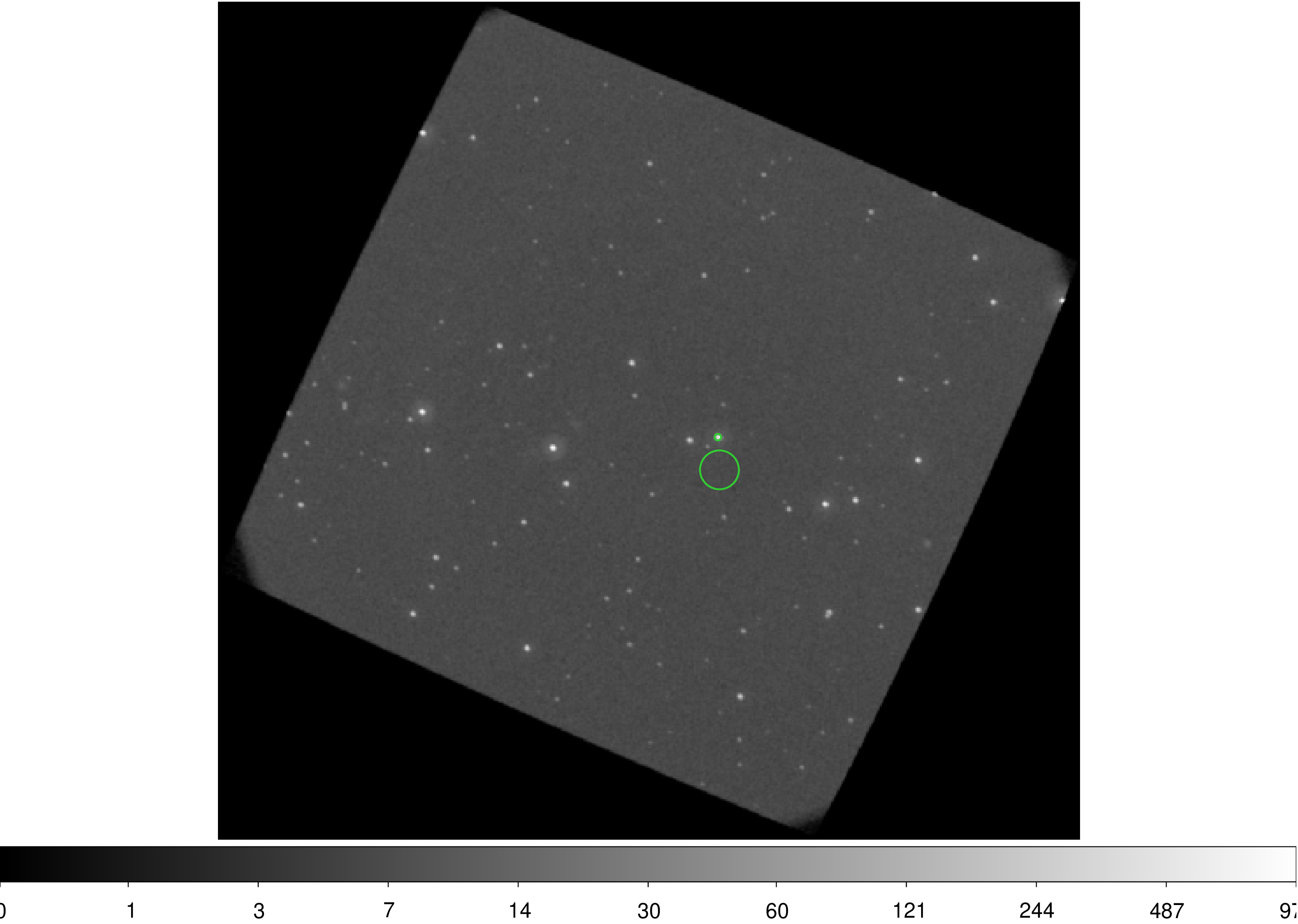}
    \includegraphics[width=0.6\linewidth]{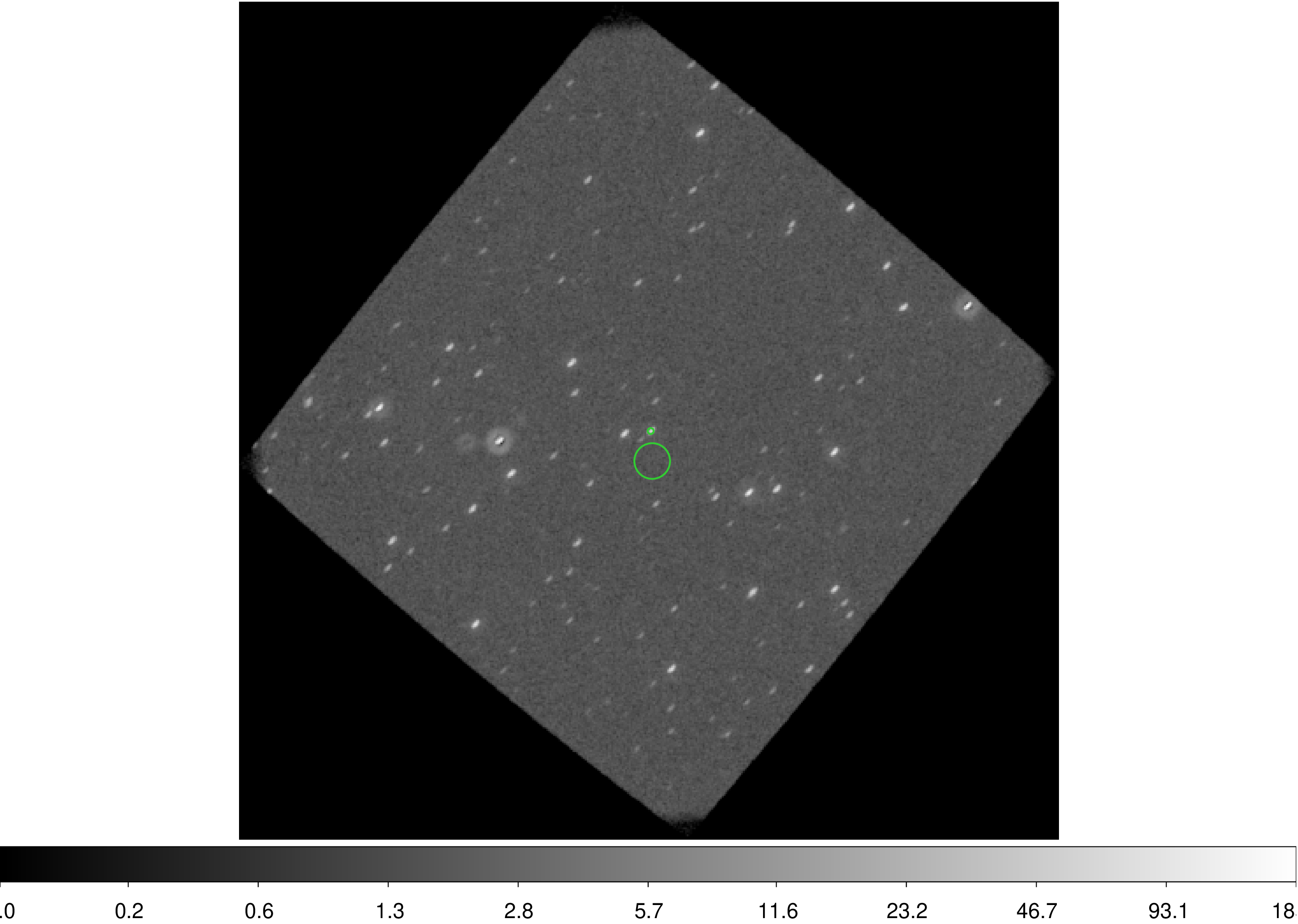}
    \includegraphics[width=0.6\linewidth]{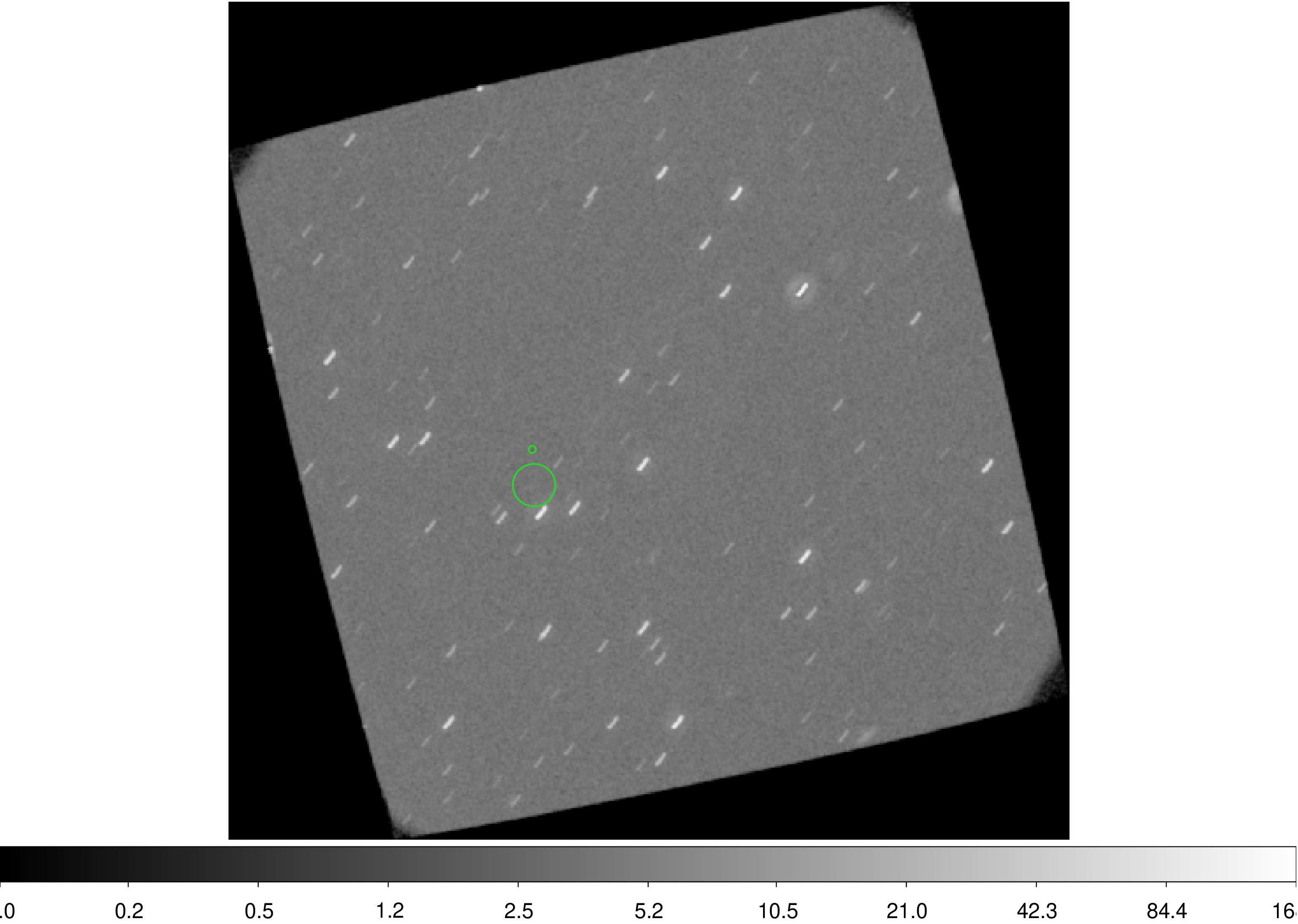}
    \caption{Examples of PNG image outputs after running the second step of the UVOT pipeline (Section~\ref{sec:data-analysis}). \textit{Top:} Image without visible issues, corresponding to a good observation. \textit{Middle}: Elongated PSF in the images, corresponding to an observation that should be discarded. \textit{Bottom}: Very deformed PSF and loss of tracking, briging the source completely out of the source region, corresponding to an observation that should be discarded.}
    \label{fig:uvot_issues}
\end{figure*}

The main issue a user of the UVOT pipeline can expect to run into are `streaks', rather than point sources, in the images (see Figure~\ref{fig:uvot_issues}). These occur when an exposure is taken as the drifts, causing a loss of appropriate tracking. There is no way of correcting for this issue, so the user should either: 1) delete the UVOT folder from within the OBSID (or the whole OBSID folder, depending on whether one is interested in running the XRT pipeline as well\footnote{We note that a small tracking error can affect UVOT observations while leaving the XRT observation usable, due to the large difference in resolution for the two telescopes.}) or, 2) run the pipeline to its end, and delete the lines corresponding to the problematic OBSID in the final file (\texttt{uvot\_mag\_flux\_all\_epochs\_filters.csv}, see Section~\ref{sec:fluxes}).

Another potential problem may be the presence of a fainter target within the chosen background region. Since the user chooses their background image based on a random exposure, it is possible that in deeper exposures and/or different filters a previously-unseen sources appears within the background region. In such cases, it is best to choose a different background region and rerun the whole analysis.

\subsection{XRT-PC}

\begin{figure*}
    \centering
    \includegraphics[width=0.45\linewidth]{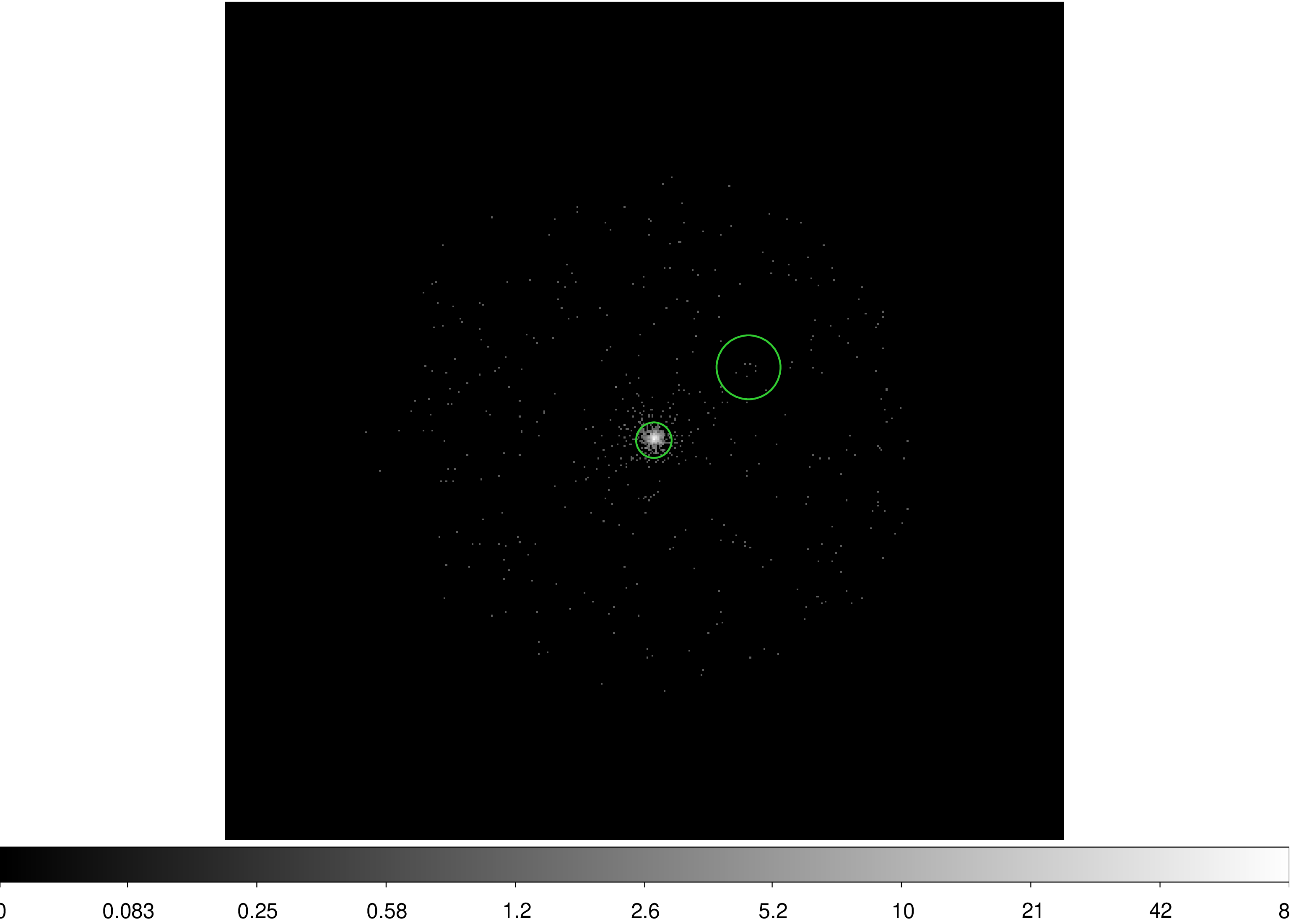}
    \includegraphics[width=0.45\linewidth]{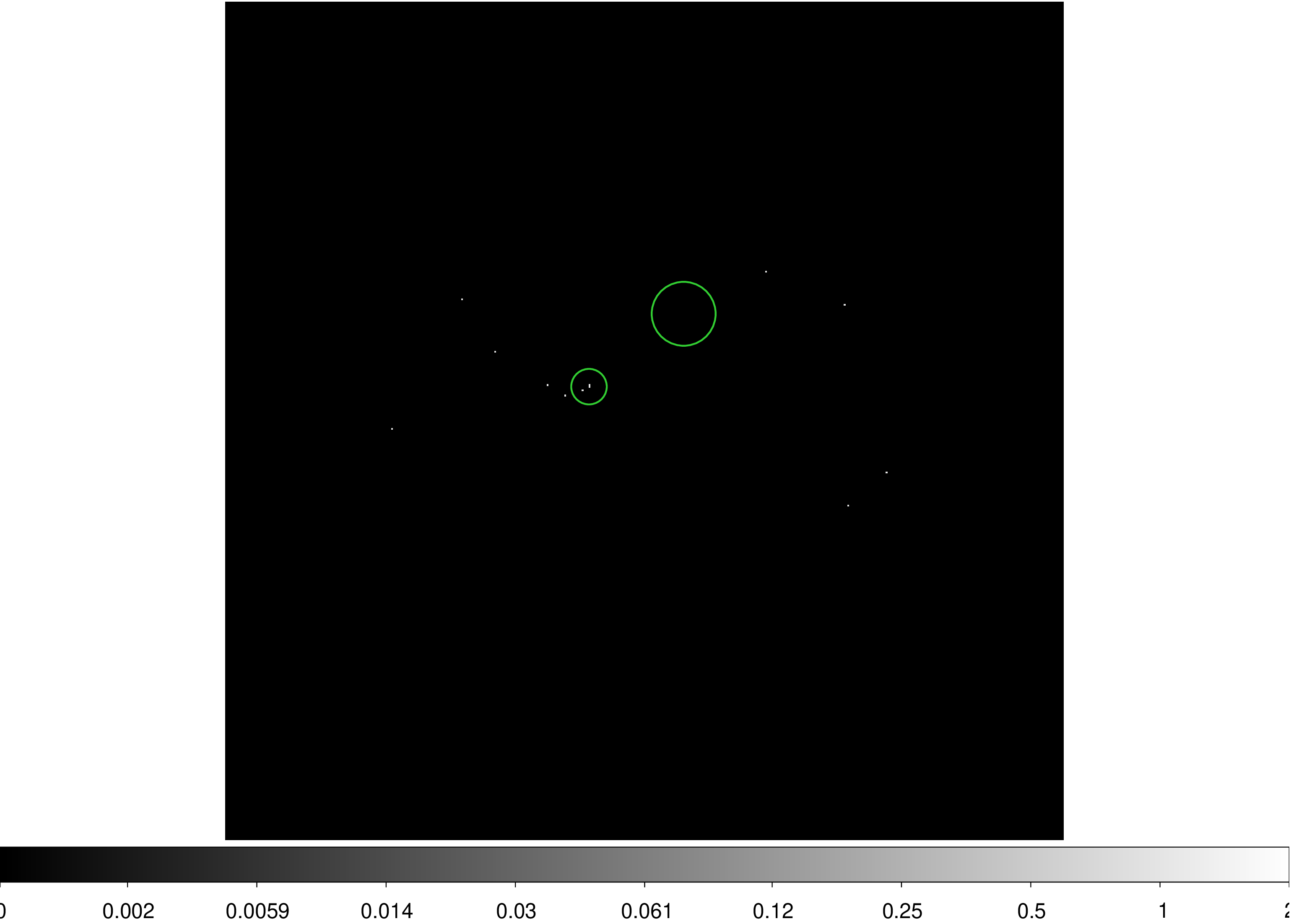}
    \includegraphics[width=0.45\linewidth]{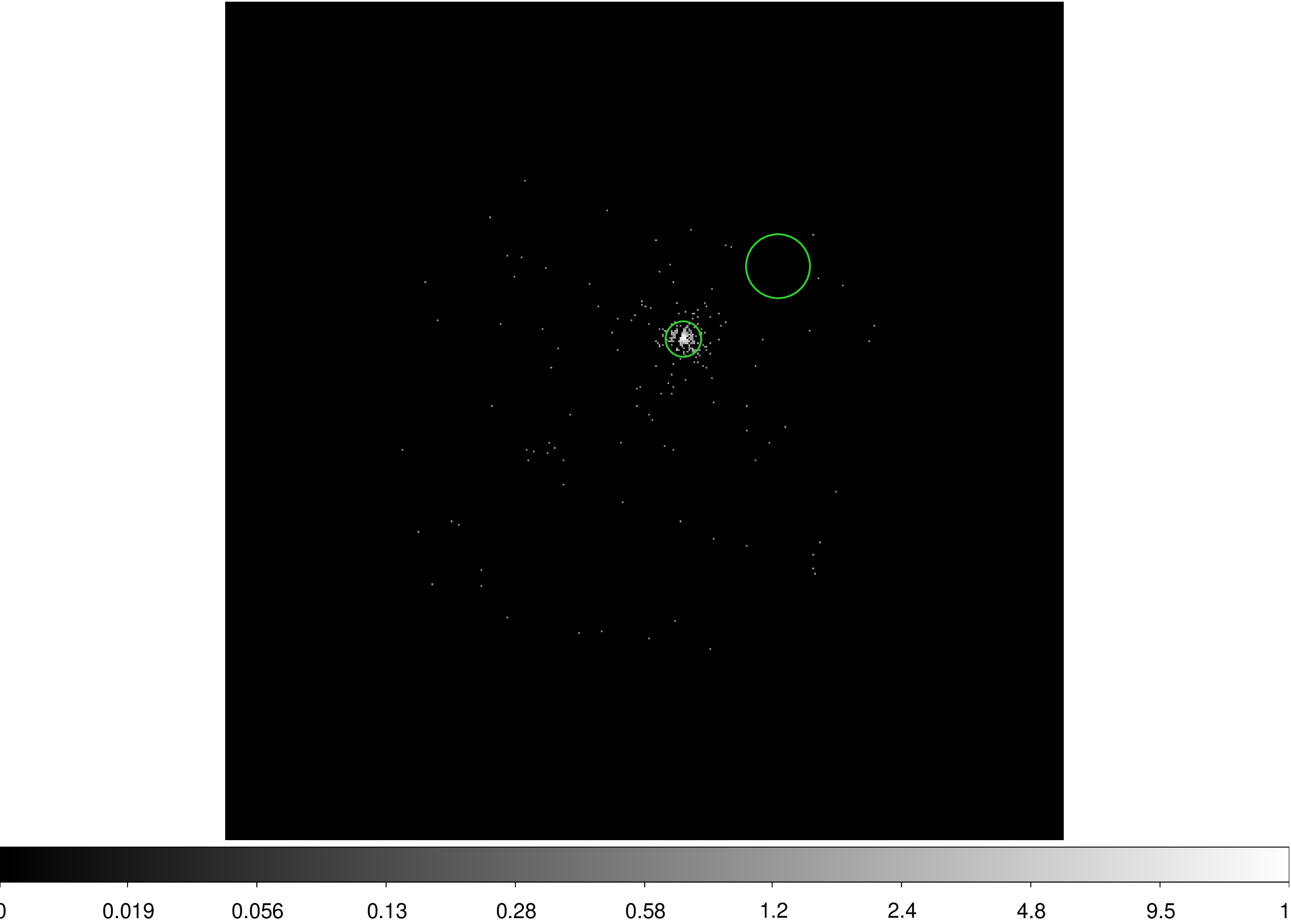}
    \includegraphics[width=0.45\linewidth]{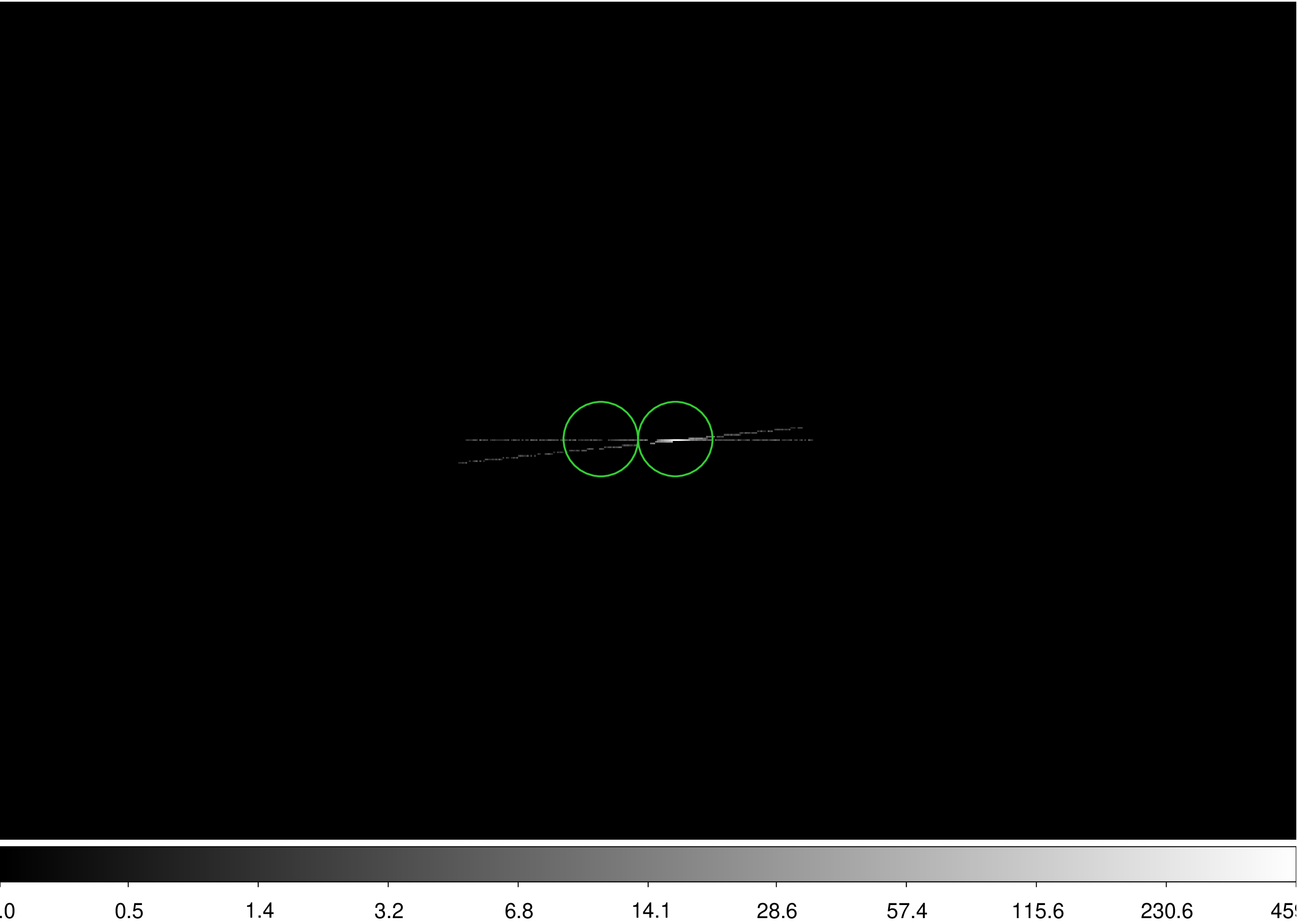}    
    \caption{Examples of PNG image outputs after running the second step of the XRT pipeline on PC and WT events (Section~\ref{sec:data-analysis}). \textit{Top left}: Image without visible issues, corresponding to a good observation. \textit{Top right}: Almost no photons are detected in this observation at the source location due to its very low exposure, and as such should be discarded. \textit{Bottom left}: Visible blank streak crossing the source emission, caused by ``bad'' or ``hot'' columns within the CCD. This effect is corrected for via the usage of exposure maps, and as such the observation can be safely used. \textit{Bottom right}: saturated exposure of WT recognizable by the presence of the diffraction spikes.}
    \label{fig:xrt_pc_issues}
\end{figure*}

When running the XRT pipeline for PC observations, one of the most likely issues a user may encounter is the lack (or almost lack) of detected photons within the source region. This is likely to happen for faint sources in very short exposures (see Figure~\ref{fig:xrt_pc_issues}). A complete lack of photons in the source region may cause issues with the pipeline. Moreover, observations with such a low count rate are not adequate for this pipeline, as it does not compute upper limits. As such, the XRT folder within the problematic OBSID should be removed. 

Another likely issue to arise is the presence of a `blank' or black streak crossing through the source. This area corresponds to the bad (or hot) columns which run through the CCD since May 2005, when the XRT detector was hit by a micrometeoroid. These bad pixels are always present in observations, although not easily visible unless a bright source sits on top of them. Since this is a known issue, it can be corrected for via the usage of exposure maps. As such, the observations can be safely used despite their appearance. 

A third issue could arise from the fact that in some exposures the target is close to the edge of the field, making the chosen background region sit outside the edge of the detector. The solution in this case is to either remove the XRT exposure or to change the background location and rerun the analysis. Unless this happens for a considerable number of observations, it is recommended to simply discard the affected observations.

Due to the large extraction region, it is very unlikely that a small loss of tracking causes the X-ray source to move outside the chosen source region. However, were that to happen for a single (or few) observations, they should be discarded. If it occurs with a large number of observations, the user should consider whether the source coordinates are correct.

Given the scarcity of X-ray sources in the sky, compared with optical/UV, it is also unlikely that a user will encounter a fainter source within the background region of deeper observations. However, were that to happen, one should proceed exactly as recommended for UVOT.

\subsection{XRT-WT}

The main source of issue with WT is the saturation of the slit due to the source being bright. This can be visible in the presence of diffraction spikes in the image (Figure~\ref{fig:xrt_pc_issues}). The user can still decide to calculate the flux of the target, keeping in mind it may be completely wrong. In this case, we suggest to cross check with the the PC flux values (if available) and to ensure the observations do not suffer from severe pile-up (see Section~\ref{sec:pileup}), or to simply discard the observation.

A secondary source of problems may be that the pointing of WT is not on your science target but on something else closeby. This issue is noted in the warning messages of the pipeline. In this case, we suggest to simply remove the observation. 

\subsection{Pile-up}\label{sec:pileup}

X-ray CCDs measure photon energy by detecting the current generated in a pixel after photon impact. Because of this, they are susceptible to pile-up -- i.e., the erroneous detection of multiple photons as a single one when they all impact the same pixel within the readout time. This will result in a single energy measurement, equivalent to the sum of the energies of the individual photons. Pile-up occurs only for bright sources, with a high-enough count rate to generate multiple photons within a single readout time. The exact flux threhsold for pile-up depends on the telescope and instrument considered.

For \xrt, pile-up occurs mainly in the PC mode. WT will, in general, not be affected by pile-up under count rates of 100~cts/s \citep{Romano2006}, which are extreme for the majority of user cases. For PC, however, this threshold is 0.5~cts/s, which is more commonly achieved. SAPLE lists count-rate for each XRT observation when generating its final output file, allowing the user to flag observations that are susceptible to pile-up. For the affected observations, we suggest one of the following:
\begin{itemize}
    \item[a.] If a WT observation is available under the same OBSID, use the WT observation only.
    \item[b.] If no WT observation is available, the user can manually check the extent of piple-up and, if needed, correct the affected observations following the official \xrt~ guidelines\footnote{\url{https://www.swift.ac.uk/analysis/xrt/pileup.php}}. While this affect can be corrected for, the current version of SAPLE does not implement any corrections.
\end{itemize}

Pile-up affects the shape of the spectrum, hardening it artificially, in a way that deviates from the shape of a power-law\footnote{We assume a power-law shape, as the current version of SAPLE only fits a power-law model.}. As such, the fit will return a harder photon index value, a higher flux, and will report a bad fit statistic. We do not recommend relying on SAPLE results for piled-up observations without any kind of correction. 

\section{Fluxes}\label{sec:fluxes}
\subsection{UVOT}

By default, \texttt{uvotsource} provides the user with magnitudes (in AB and Vega systems) and flux densities. The magnitudes are calibrated using the zero points \citep{Poole_2008,Breeveld_2011}: 
\begin{equation}
   m = ZPT - 2.5 \times \log_{10}(C) 
\end{equation}

where $ZPT$ are the zeroth magnitude calibrated for ``standard sources" and are stored in the CALDB files and $C$ is the count rate of the source in a specific filter. Similarly, the flux densities are calculated as: 
\begin{equation}
    F_{\lambda} = FCF \times C
\end{equation}
where $FCF$ are flux conversion factors applied directly to the count rates. 

However, the UV magnitudes (and fluxes) provided by the \texttt{uvotsource} pipeline are \textit{not} corrected by UV extinction. This is a rather non-trivial correction, and at the moment in SAPLE we rely on the correction reported for GRBs\footnote{Note that if the source's spectrum is not akin to a GRB and if the source is located in galaxies with extiction curves different than the Milky Way one, this correction is an oversimplification and may lead to incorrect results.} in \citet{Roming_2009}. Therefore, in SAPLE we use the AB magnitudes extracted with \texttt{uvotsource}, calculate the UV extinction ($A_{\lambda}$) in every filter using Equation 2 in \citet{Roming_2009}, compute the corrected AB magnitudes as: 
\begin{equation}\label{eq:uvcorr}
  m'_{\rm AB,\lambda} = m_{\rm AB,\lambda}-A_{\lambda}
\end{equation}
where $\lambda$ is the wavelength of the filter you are considering; and then compute the specific fluxes using the following: 
\begin{equation}\label{eq:fdens}
    F_{\nu} = 10^{(m'_{\rm AB, \lambda}+48.6)/-2.5} 
\end{equation}

where the units for $F_{\nu}$ are [erg/cm$^2$/s/Hz].
We note that recently, \citet{Yi_2023} provided empirical reddening and extinction coefficients for the \uvot~ passbands. Integration of those extinction values are under implementation in SAPLE.  

The final product of the SAPLE UVOT pipeline is a CVS file (\texttt{uvot\_mag\_flux\_all\_epochs\_filters.csv}) that contains information about the filter, time of the observation, observation number, AB magnitudes and uncertainties (statistic and systematic) and fluxes from the \texttt{uvotsource} routine (not corrected by UV extinction), plus the AB absorption corrected magnitudes and flux densities (and uncertainties) calculated via Equations~\ref{eq:uvcorr}-\ref{eq:fdens}. Table~\ref{tab:uvot_output} contains all the information about the output CSV file columns. 

\begin{deluxetable*}{ccl}
\tablecaption{Column description UVOT spectral products in \texttt{uvot\_mag\_flux\_all\_epochs\_filters.csv} \label{tab:uvot_output}}
\tablehead{
\colhead{Column Name} & \colhead{Units} & \colhead{Description}
}
\startdata
obsid &  & Observation ID number \\
filt &  & UVOT filter (v, b, u, uvw1, uvm2, uvw2) \\
tstart[MET] & MET & Start time of the observation in MET \\
tstop[MET]  & MET & End time of the observation in MET \\
tstart[MJD] & MJD & Start time of the observation in MJD \\
tstop[MJD]  & MJD & End time of the observation in MJD \\
nu[hz]      & Hz  & Frequency of the filter \\
wave[AA]    & \AA & Wavelength of the filter \\
\midrule
ab\_mag            &        & Source AB magnitude (from \texttt{uvotsource}) \\
ab\_mag\_err       &        & Source AB magnitude total error (from \texttt{uvotsource}) \\
ab\_mag\_err\_stat &        & Source AB magnitude statistical error (from \texttt{uvotsource}) \\
ab\_mag\_err\_syst &        & Source AB magnitude systematic error (from \texttt{uvotsource}) \\
ab\_mag\_ext\_corr &        & Source AB magnitude corrected for extinction (SAPLE, Eq.~\ref{eq:uvcorr}) \\
\midrule
sflux\_hz\_uvot[erg/cm2/s/Hz] & erg\,cm$^{-2}$\,s$^{-1}$\,Hz$^{-1}$ & Source flux density (from \texttt{uvotsource})\tablenotemark{$\ddagger$} \\
sflux\_hz\_err\_stat\_uvot[erg/cm2/s/Hz] & erg\,cm$^{-2}$\,s$^{-1}$\,Hz$^{-1}$ & Source flux density statistical error (from \texttt{uvotsource})\tablenotemark{$\dagger$} \\
sflux\_hz\_err\_syst\_uvot[erg/cm2/s/Hz] & erg\,cm$^{-2}$\,s$^{-1}$\,Hz$^{-1}$ & Source flux density systematic error (from \texttt{uvotsource})\tablenotemark{$\dagger$} \\
sflux\_hz\_bkg\_uvot[erg/cm2/s/Hz] & erg\,cm$^{-2}$\,s$^{-1}$\,Hz$^{-1}$ & Background flux density statistical error (from \texttt{uvotsource})\tablenotemark{$\dagger$} \\
\midrule
sflux\_hz\_saple[erg/cm2/s/Hz] & erg\,cm$^{-2}$\,s$^{-1}$\,Hz$^{-1}$ & Source flux density in frequency space (from SAPLE, Eq.~\ref{eq:fdens})\tablenotemark{$\ddagger$}\\
sflux\_hz\_err\_d\_saple[erg/cm2/s/Hz] & erg\,cm$^{-2}$\,s$^{-1}$\,Hz$^{-1}$ & Lower flux density uncertainty (frequency, from SAPLE, Eq.~\ref{eq:fdens})\tablenotemark{$\ddagger$} \\
sflux\_hz\_err\_u\_saple[erg/cm2/s/Hz] & erg\,cm$^{-2}$\,s$^{-1}$\,Hz$^{-1}$ & Upper flux density uncertainty (frequency, from SAPLE, Eq.~\ref{eq:fdens})\tablenotemark{$\ddagger$} \\
sflux\_wave\_saple[erg/cm2/s/AA] & erg\,cm$^{-2}$\,s$^{-1}$\,\AA$^{-1}$ & Source flux density in wavelength space (from SAPLE, Eq.~\ref{eq:fdens})\tablenotemark{$\ddagger$} \\
sflux\_wave\_err\_d\_saple[erg/cm2/s/AA] & erg\,cm$^{-2}$\,s$^{-1}$\,\AA$^{-1}$ & Lower flux density uncertainty (wavelength, from SAPLE, Eq.~\ref{eq:fdens})\tablenotemark{$\ddagger$}  \\
sflux\_wave\_err\_u\_saple[erg/cm2/s/AA] & erg\,cm$^{-2}$\,s$^{-1}$\,\AA$^{-1}$ & Upper flux density uncertainty (wavelength, from SAPLE, Eq.~\ref{eq:fdens})\tablenotemark{$\ddagger$}  \\
upper\_limit &  & Reported magnitude (flux) is an upper (lower) limit: 0-NO; 1-YES
\enddata
\tablenotetext{$\dagger$}{Computed without correcting magnitudes by Galactic extinction.}
\tablenotetext{$\ddagger$}{Computed after correcting magnitudes by Galactic extinction.}
\end{deluxetable*}

\subsection{XRT}
For both the WT and PC spectral files, for every observation the pipeline does the following:
\begin{itemize}
    \item[a.] Creates an Ancillary Response File (ARF);
    \item[b.] Using \texttt{grppha}, it associates the background spectrum, ARF and RMF to the source spectrum;
    \item[c.] Using \texttt{ftgrppha}, it groups the source spectrum using optimal binning;
    \item[d.] With pyXSPEC, it fits the spectrum with an absorbed redshifted powerlaw (\texttt{TBAbs*zpowerlaw}), with absorption fixed at the Galactic level and abundance set to the \citet{Wilms_2000} value; by default, the fit is performed using Cash statistics (\texttt{cstat}) over the $0.3-10\,\rm keV$ band. 
\end{itemize}

The final product of the pipeline is a CVS file (\texttt{params\_xspec.csv}) that contains information about the exposure, time of the observation, observation number, instrument of the observation, count rate in the $0.3-10\,\rm keV$ band, powerlaw index and uncertainty at the 90\% confidence level (C.L.), absorbed powerlaw flux and uncertainty at the 90\% C.L. and unabsorbed powerlaw flux. Table~\ref{tab:xray_output} contains all the information about the output CSV file columns.

\begin{deluxetable*}{ccl}
\tablecaption{Column description of X-ray spectral products in \texttt{params\_xspec.csv}\label{tab:xray_output}}
\tablehead{
\colhead{Column Name} & \colhead{Units} & \colhead{Description}
}
\startdata
inst &  & XRT instrument mode, WT or PC \\
obs\_id &  &  Observation ID number \\
exposure\_time[s] & s & Observation exposure time \\
start\_date & YYYY-MM-DD &  Start date of the observation \\
start\_date[mjd] & MJD &  Start date of the observation in MJD \\
count/s & cts s$^{-1}$ & Source counts per second in the $0.3-10\,\rm keV$ band \\
pl\_idx &  &  Source power-law index in the $0.3-10\,\rm keV$ band \\
pl\_idx\_l &  & Source power-law index lower value (90\% CL) \\
pl\_idx\_h &  & Source power-law index upper value (90\% CL) \\
flux\_abs & erg\,cm$^{-2}$\,s$^{-1}$ & Source \textit{absorbed} power-law flux in the $0.3-10\,\rm keV$ band \\
flux\_abs\_err\_l & erg\,cm$^{-2}$\,s$^{-1}$ & Source \textit{absorbed} power-law flux lower uncertainty value (90\% CL)\\
flux\_abs\_err\_h & erg\,cm$^{-2}$\,s$^{-1}$ & Source \textit{absorbed} power-law flux upper uncertainty value (90\% CL) \\
flux\_unabs & erg\,cm$^{-2}$\,s$^{-1}$ & Source \textit{unabsorbed} power-law flux in the $0.3-10\,\rm keV$ band \\
cstat &  & Cash Statistic \\
dof &  & Degrees of freedom
\enddata
\end{deluxetable*}

\section{Pipeline products \& Result visualization}\label{sec:products}
The full suite of codes that constitute SAPLE is available at the GitHub repository \citet{marcotulli2026}.
Given that SAPLE is a semi-automated pipeline, the user needs to follow a step-by-step procedure to obtain the data products. In the GitHub, we provide a separate \texttt{README} file for both the \uvot~ and the \xrt~ codes that we encourage the user to read carefully before proceding with the analysis. Any errors should occurr, please contact the authors or open an issue on GitHub. 

\begin{figure*}
    \centering
    \includegraphics[width=0.8\linewidth]{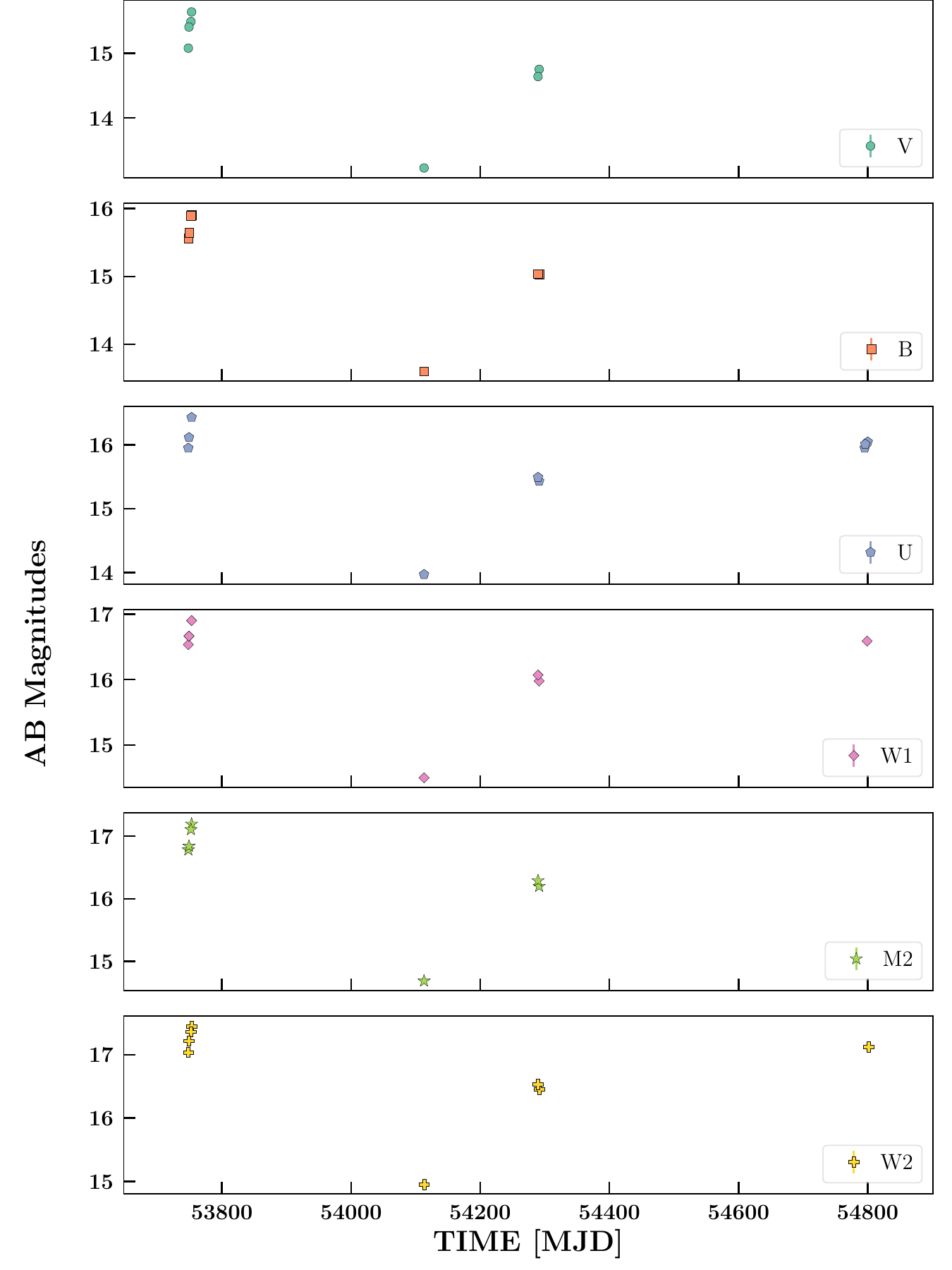}
    \caption{\uvot~ extinction corrected magnitudes in the 6 filters (U, V, B, W1, M2, W2) light-curves for the blazar 3C 279 (obsid: 00030867001, 00030867010, 00030867011, 00035019001, 00035019002, 00035019004, 00035019005, 00035019007, 00035019009, 00035019010, 00035019011). The magnitude points that are separated in time by only a few days look as if they were overlapping, and not all the pointings had data in all filters. Upper limits have been excluded from this plot.}
    \label{fig:uvot_lc}
\end{figure*}

\begin{figure*}
    \centering
    \includegraphics[width=0.8\linewidth]{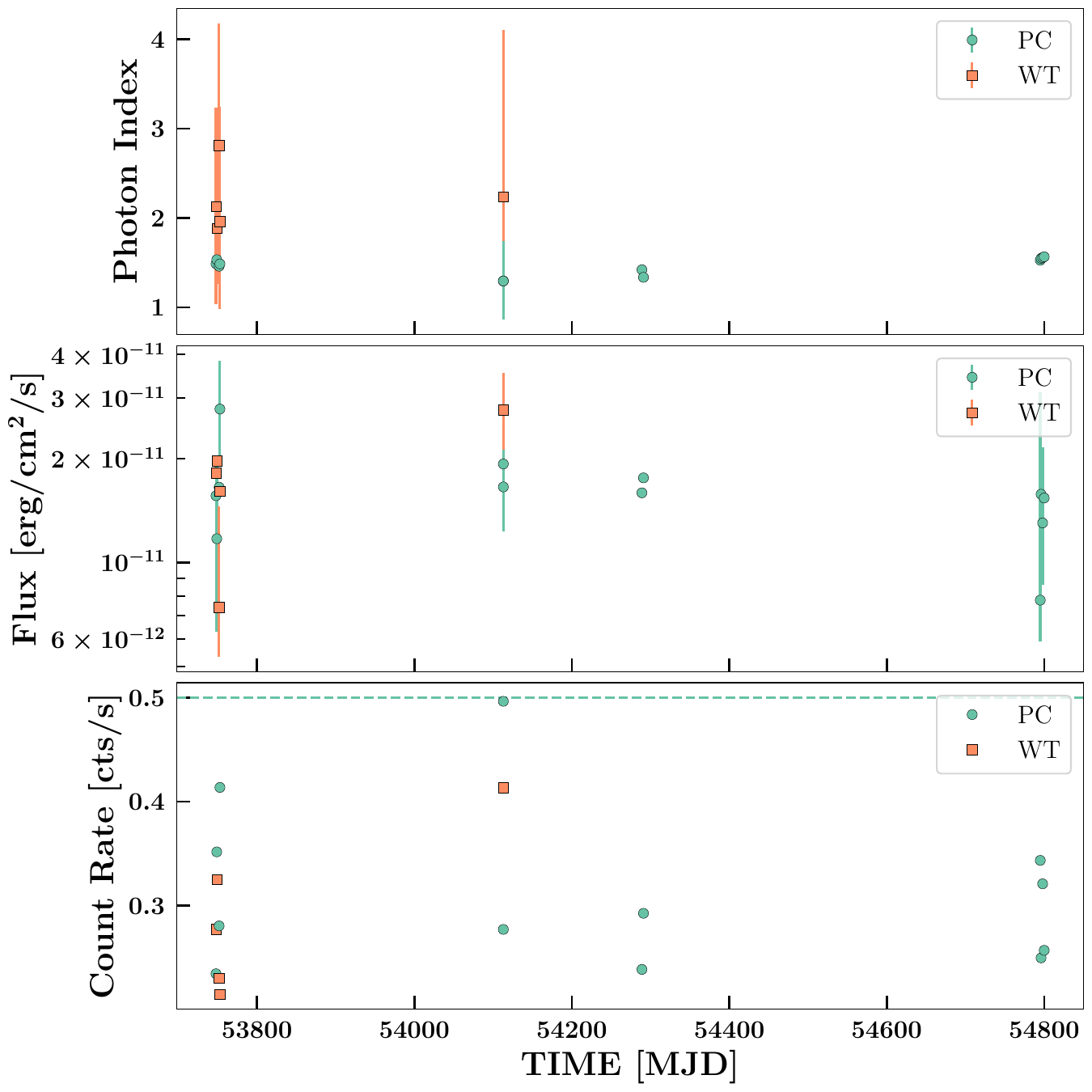}
    \caption{\xrt~ PC (cyan points) and WT (orange squares) photon index (top), unabsorbed power-law flux (middle), and count rates (bottom) ligthcurve in the $0.3-10\,\rm keV$ band for the blazar 3C 279 (obsid: 00030867001, 00030867010, 00030867011, 00035019001, 00035019002, 00035019004, 00035019005, 00035019007, 00035019009, 00035019010, 00035019011). Rows of zero flux values have been excluded from this plot. All the count rates are below the PC pileup threshold value of 0.5 cts/s (cyan dashed line), indicating that the observations do not suffer from pileup (for WT this threshold is 100 cts/s).}
    \label{fig:xrt_lc}
\end{figure*}

After running the complete pipeline on all the observations, the user will have two separate CSV files: \texttt{uvot\_mag\_flux\_all\_epochs\_filters.csv} (for \uvot, see Table~\ref{tab:uvot_output}); \texttt{params\_xspec.csv} (for \xrt, see Table~\ref{tab:xray_output}). One way the user can visualize the results is by plotting the lightcurves. In Figure~\ref{fig:uvot_lc} and \ref{fig:xrt_lc}, we provide some examples for both the UVOT and XRT lightcurves on eleven observations for 3C 279\footnote{Obsid: 00030867001, 00030867010, 00030867011, 00035019001, 00035019002, 00035019004, 00035019005, 00035019007, 00035019009, 00035019010, 00035019011}, a blazar source located at $z=0.53$. For UVOT, we show the extinction corrected AB magnitude lightcurves for all 6 UVOT filters (U, B, V, W1, M2, W2). For XRT (WT and PC), we show the photon index, the absorption corrected flux for the power-law fit, and the count rates of the source in the $0.3-10\,\rm keV$ band. The count rates for both PC and WT of the target are all below the PC threshold for pileup (0.5 cts/s), hence no further corrections is needed to extract the source's spectra.  

In its beta version, this pipeline has already been used in scientific publications \citep[][]{Penil_2024a, Penil_2024b} and in its v.1.0 in  \citet{Penil_2026}. Considering the upcoming, but not immediate, launch of future UV mission (such as UVEX and ULTRASAT, \citealp{uvex_2021,ultrasat_2022}) and X-ray missions (such as {\it NewAthena}, \citealp{Cruise_2025}), we encourage the community to make use of the wealth of data already available in the archive to explore the fast and transient UV and X-ray sky.

\section{Future development on the X-ray analysis}\label{sec:future}
SAPLE X-ray fitting analysis is currently limited to a `simple' redshifted power-law fit. In the future developments of the pipeline, we plan to implement more complex shapes to be fit to every observations (e.g.~a broken-powerlaw, a log-parabola, etc.) with the possibility of performing model comparison. We will also implement the fit with the Bayesian X-ray software (BXA, \citealp[][]{Buchner_2014}) and use the corresponding Bayes factors to perform accurate model comparison \citep{Buchner_2023}.
The X-ray fitting code uses by default the spectra binned via optimal binning and has a fixed energy band (0.3-10 keV). In the future, we plan to allow for user input on choice of binning and fitted energy band. We also encourage any user to give us suggestions on what they would like to see implemented in SAPLE.

\section*{Acknowledgments}
LM and NT thank the \swift~team for promptly answering all the questions regarding the \xrt~and \uvot~analysis we asked while working with SAPLE. LM thanks  Mr. Felix Fischer for fruitful discussion on the UV absorption. LM and NT would like to thank Dr.~Dr.~Pe{\~n}il for providing the first dataset that required the development of SAPLE.
LM acknowledges that this work was supported by the Initiative and Networking Fund of the Helmholtz Association under the Helmholtz Investigator Groups Programme, call 2025 (VH-NG-21-01). LM acknowledges support from DESY (Zeuthen, Germany), a member of the Helmholtz Association HGF.
\\

{\it Facilities:}\uvot, \xrt


\bibliography{xrt_uvot}{}

@INPROCEEDINGS{xspec_1996,
       author = {{Arnaud}, K.~A.},
        title = "{XSPEC: The First Ten Years}",
    booktitle = {Astronomical Data Analysis Software and Systems V},
         year = 1996,
       editor = {{Jacoby}, George H. and {Barnes}, Jeannette},
       series = {Astronomical Society of the Pacific Conference Series},
       volume = {101},
        month = jan,
        pages = {17},
       adsurl = {https://ui.adsabs.harvard.edu/abs/1996ASPC..101...17A}
}

@ARTICLE{Penil_2026,
       author = {{Pe{\~n}il}, P. and {Torres-Alb{\`a}}, N. and {Marcotulli}, L. and {Dom{\'\i}nguez}, A. and {Ajello}, M. and {Rico}, A. and {Buson}, S. and {Adhikari}, S.},
        title = "{Testing X-ray Periodicity and Long-Term Trend in PG 1553+113 via Targeted Swift-XRT Monitoring}",
      journal = {arXiv e-prints},
         year = 2026,
        month = apr,
          eid = {arXiv:2604.05905},
        pages = {arXiv:2604.05905},
          doi = {10.48550/arXiv.2604.05905},
archivePrefix = {arXiv},
       eprint = {2604.05905},
 primaryClass = {astro-ph.HE},
       adsurl = {https://ui.adsabs.harvard.edu/abs/2026arXiv260405905P}
}

@INCOLLECTION{Buchner_2023,
       author = {{Buchner}, Johannes and {Boorman}, Peter},
        title = "{Statistical Aspects of X-ray Spectral Analysis}",
    booktitle = {Handbook of X-ray and Gamma-ray Astrophysics},
    publisher = {Springer Singapore},
         year = 2023,
          eid = {150},
        pages = {150},
          doi = {10.1007/978-981-16-4544-0_175-1},
       adsurl = {https://ui.adsabs.harvard.edu/abs/2023hxga.book..150B}
}

@ARTICLE{Buchner_2014,
       author = {{Buchner}, J. and {Georgakakis}, A. and {Nandra}, K. and {Hsu}, L. and {Rangel}, C. and {Brightman}, M. and {Merloni}, A. and {Salvato}, M. and {Donley}, J. and {Kocevski}, D.},
        title = "{X-ray spectral modelling of the AGN obscuring region in the CDFS: Bayesian model selection and catalogue}",
      journal = {\aap},
         year = 2014,
        month = apr,
       volume = {564},
          eid = {A125},
        pages = {A125},
          doi = {10.1051/0004-6361/201322971},
archivePrefix = {arXiv},
       eprint = {1402.0004},
 primaryClass = {astro-ph.HE},
       adsurl = {https://ui.adsabs.harvard.edu/abs/2014A&A...564A.125B}
}

@ARTICLE{Cruise_2025,
       author = {{Cruise}, Mike and {Guainazzi}, Matteo and {Aird}, James and {Carrera}, Francisco J. and {Costantini}, Elisa and {Corrales}, Lia and {Dauser}, Thomas and {Eckert}, Dominique and {Gastaldello}, Fabio and {Matsumoto}, Hironori and {Osten}, Rachel and {Petrucci}, Pierre-Olivier and {Porquet}, Delphine and {Pratt}, Gabriel W. and {Rea}, Nanda and {Reiprich}, Thomas H. and {Simionescu}, Aurora and {Spiga}, Daniele and {Troja}, Eleonora},
        title = "{The NewAthena mission concept in the context of the next decade of X-ray astronomy}",
      journal = {Nature Astronomy},
         year = 2025,
        month = jan,
       volume = {9},
        pages = {36-44},
          doi = {10.1038/s41550-024-02416-3},
archivePrefix = {arXiv},
       eprint = {2501.03100},
 primaryClass = {astro-ph.IM},
       adsurl = {https://ui.adsabs.harvard.edu/abs/2025NatAs...9...36C}
}

@ARTICLE{uvex_2021,
       author = {{Kulkarni}, S.~R. and {Harrison}, Fiona A. and {Grefenstette}, Brian W. and {Earnshaw}, Hannah P. and {Andreoni}, Igor and {Berg}, Danielle A. and {Bloom}, Joshua S. and {Cenko}, S. Bradley and {Chornock}, Ryan and {Christiansen}, Jessie L. and {Coughlin}, Michael W. and {Wuollet Criswell}, Alexander and {Darvish}, Behnam and {Das}, Kaustav K. and {De}, Kishalay and {Dessart}, Luc and {Dixon}, Don and {Dorsman}, Bas and {El-Badry}, Kareem and {Evans}, Christopher and {Ford}, K.~E. Saavik and {Fremling}, Christoffer and {Gansicke}, Boris T. and {Gezari}, Suvi and {Goetberg}, Y. and {Green}, Gregory M. and {Graham}, Matthew J. and {Heida}, Marianne and {Ho}, Anna Y.~Q. and {Jaodand}, Amruta D. and {Johns-Krull}, Christopher M. and {Kasliwal}, Mansi M. and {Lazzarini}, Margaret and {Lu}, Wenbin and {Margutti}, Raffaella and {Martin}, D. Christopher and {Masters}, Daniel Charles and {McKernan}, Barry and {Naze}, Yael and {Nissanke}, Samaya M. and {Parazin}, B. and {Perley}, Daniel A. and {Phinney}, E. Sterl and {Piro}, Anthony L. and {Raaijmakers}, G. and {Rauw}, Gregor and {Rodriguez}, Antonio C. and {Sana}, Hugues and {Senchyna}, Peter and {Singer}, Leo P. and {Spake}, Jessica J. and {Stassun}, Keivan G. and {Stern}, Daniel and {Teplitz}, Harry I. and {Weisz}, Daniel R. and {Yao}, Yuhan},
        title = "{Science with the Ultraviolet Explorer (UVEX)}",
      journal = {arXiv e-prints},
         year = 2021,
        month = nov,
          eid = {arXiv:2111.15608},
        pages = {arXiv:2111.15608},
          doi = {10.48550/arXiv.2111.15608},
archivePrefix = {arXiv},
       eprint = {2111.15608},
 primaryClass = {astro-ph.GA},
       adsurl = {https://ui.adsabs.harvard.edu/abs/2021arXiv211115608K}
}

@INPROCEEDINGS{ultrasat_2022,
       author = {{Ben-Ami}, Sagi and {Shvartzvald}, Yossi and {Waxman}, Eli and {Netzer}, Udi and {Yaniv}, Yoram and {Algranatti}, Viktor M. and {Gal-Yam}, Avishay and {Lapid}, Ofer and {Ofek}, Eran and {Topaz}, Jeremy and {Arcavi}, Iair and {Asif}, Arooj and {Azaria}, Shlomi and {Bahalul}, Eran and {Barschke}, Merlin F. and {Bastian-Querner}, Benjamin and {Berge}, David and {Berlea}, Vlad D. and {Buehler}, Rolf and {Dittmar}, Louise and {Gelman}, Anatoly and {Giavitto}, Gianluca and {Guttman}, Or and {Haces Crespo}, Juan M. and {Heilbrunn}, Daniel and {Kachergincky}, Arik and {Kaipachery}, Nirmal and {Kowalski}, Marek and {Kulkarni}, Shrinivasrao R. and {Kumar}, Shashank and {K{\"u}sters}, Daniel and {Liran}, Tuvia and {Miron-Salomon}, Yonit and {Mor}, Zohar and {Nir}, Aharon and {Nitzan}, Gadi and {Philipp}, Sebastian and {Porelli}, Andrea and {Sagiv}, Ilan and {Schliwinski}, Julian and {Sprecher}, Tuvia and {De Simone}, Nicola and {Stern}, Nir and {Stone}, Nicholas C. and {Trakhtenbrot}, Benny and {Vasilev}, Mikhail and {Watson}, Jason J. and {Zappon}, Francesco},
        title = "{The scientific payload of the Ultraviolet Transient Astronomy Satellite (ULTRASAT)}",
    booktitle = {Space Telescopes and Instrumentation 2022: Ultraviolet to Gamma Ray},
         year = 2022,
       editor = {{den Herder}, Jan-Willem A. and {Nikzad}, Shouleh and {Nakazawa}, Kazuhiro},
       series = {SPIE},
       volume = {12181},
        month = aug,
          eid = {1218105},
        pages = {1218105},
          doi = {10.1117/12.2629850},
archivePrefix = {arXiv},
       eprint = {2208.00159},
 primaryClass = {astro-ph.IM},
       adsurl = {https://ui.adsabs.harvard.edu/abs/2022SPIE12181E..05B}
}

@ARTICLE{Wilms_2000,
       author = {{Wilms}, J. and {Allen}, A. and {McCray}, R.},
        title = "{On the Absorption of X-Rays in the Interstellar Medium}",
      journal = {\apj},
         year = 2000,
        month = oct,
       volume = {542},
       number = {2},
        pages = {914-924},
          doi = {10.1086/317016},
archivePrefix = {arXiv},
       eprint = {astro-ph/0008425},
 primaryClass = {astro-ph},
       adsurl = {https://ui.adsabs.harvard.edu/abs/2000ApJ...542..914W}
}

@INPROCEEDINGS{Breeveld_2011,
       author = {{Breeveld}, A.~A. and {Landsman}, W. and {Holland}, S.~T. and {Roming}, P. and {Kuin}, N.~P.~M. and {Page}, M.~J.},
        title = "{An Updated Ultraviolet Calibration for the Swift/UVOT}",
    booktitle = {Gamma Ray Bursts 2010},
         year = 2011,
       editor = {{McEnery}, J.~E. and {Racusin}, J.~L. and {Gehrels}, N.},
       series = {AIP},
       volume = {1358},
        month = aug,
    publisher = {AIP},
        pages = {373-376},
          doi = {10.1063/1.3621807},
archivePrefix = {arXiv},
       eprint = {1102.4717},
 primaryClass = {astro-ph.IM},
       adsurl = {https://ui.adsabs.harvard.edu/abs/2011AIPC.1358..373B}
}

@ARTICLE{Poole_2008,
       author = {{Poole}, T.~S. and {Breeveld}, A.~A. and {Page}, M.~J. and {Landsman}, W. and {Holland}, S.~T. and {Roming}, P. and {Kuin}, N.~P.~M. and {Brown}, P.~J. and {Gronwall}, C. and {Hunsberger}, S. and {Koch}, S. and {Mason}, K.~O. and {Schady}, P. and {vanden Berk}, D. and {Blustin}, A.~J. and {Boyd}, P. and {Broos}, P. and {Carter}, M. and {Chester}, M.~M. and {Cucchiara}, A. and {Hancock}, B. and {Huckle}, H. and {Immler}, S. and {Ivanushkina}, M. and {Kennedy}, T. and {Marshall}, F. and {Morgan}, A. and {Pandey}, S.~B. and {de Pasquale}, M. and {Smith}, P.~J. and {Still}, M.},
        title = "{Photometric calibration of the Swift ultraviolet/optical telescope}",
      journal = {\mnras},
         year = 2008,
        month = jan,
       volume = {383},
       number = {2},
        pages = {627-645},
          doi = {10.1111/j.1365-2966.2007.12563.x},
archivePrefix = {arXiv},
       eprint = {0708.2259},
 primaryClass = {astro-ph},
       adsurl = {https://ui.adsabs.harvard.edu/abs/2008MNRAS.383..627P}
}

@misc{Gordon_2021,
       author = {{Gordon}, Craig and {Arnaud}, Keith},
        title = "{PyXspec: Python interface to XSPEC spectral-fitting program}",
 howpublished = {Astrophysics Source Code Library, record ascl:2101.014},
         year = 2021,
        month = jan,
          eid = {ascl:2101.014},
archivePrefix = {ascl},
       eprint = {2101.014},
       adsurl = {https://ui.adsabs.harvard.edu/abs/2021ascl.soft01014G}
}

@misc{heasoft_2014,
       author = {{Heasarc}},
        title = "{HEAsoft: Unified Release of FTOOLS and XANADU}",
 howpublished = {Astrophysics Source Code Library, record ascl:1408.004},
         year = 2014,
        month = aug,
          eid = {ascl:1408.004},
archivePrefix = {ascl},
       eprint = {1408.004},
       adsurl = {https://ui.adsabs.harvard.edu/abs/2014ascl.soft08004N}
}

@misc{marcotulli2026,
  author = {Lea Marcotulli and Nuria Torres-Alb{\`a}},
  title = {SAPLE - Swift Analysis Pipeline for Lightcurve Extraction},
  version = {v1.0.0},
  year = {2026},
  url = {https://github.com/leamarcotulli/saple/},
  publisher = {GitHub}
}

@ARTICLE{Penil_2024b,
       author = {{Pe{\~n}il}, P. and {Otero-Santos}, J. and {Ajello}, M. and {Buson}, S. and {Dom{\'\i}nguez}, A. and {Marcotulli}, L. and {Torres-Alb{\`a}}, N. and {Becerra Gonz{\'a}lez}, J. and {Acosta-Pulido}, J.~A.},
        title = "{Multiwavelength variability analysis of Fermi-LAT blazars}",
      journal = {\mnras},
         year = 2024,
        month = apr,
       volume = {529},
       number = {2},
        pages = {1365-1385},
          doi = {10.1093/mnras/stae594},
archivePrefix = {arXiv},
       eprint = {2403.01520},
 primaryClass = {astro-ph.HE},
       adsurl = {https://ui.adsabs.harvard.edu/abs/2024MNRAS.529.1365P}
}

@ARTICLE{Penil_2024a,
       author = {{Pe{\~n}il}, P. and {Westernacher-Schneider}, J.~R. and {Ajello}, M. and {Dom{\'\i}nguez}, A. and {Buson}, S. and {Otero-Santos}, J. and {Marcotulli}, L. and {Torres-Alb{\`a}}, N. and {Zrake}, J.},
        title = "{Multiwavelength analysis of Fermi-LAT blazars with high-significance periodicity: detection of a long-term rising emission in PG 1553+113}",
      journal = {\mnras},
         year = 2024,
        month = feb,
       volume = {527},
       number = {4},
        pages = {10168-10184},
          doi = {10.1093/mnras/stad3246},
archivePrefix = {arXiv},
       eprint = {2310.12754},
 primaryClass = {astro-ph.HE},
       adsurl = {https://ui.adsabs.harvard.edu/abs/2024MNRAS.52710168P}
}

@ARTICLE{Tohuvavohu_2024,
       author = {{Tohuvavohu}, Aaron and {Kennea}, Jamie A. and {Roberts}, Christopher J. and {DeLaunay}, James and {Ronchini}, Samuele and {Cenko}, S. Bradley and {Ewing}, Becca and {Magee}, Ryan and {Messick}, Cody and {Sachdev}, Surabhi and {Singer}, Leo P.},
        title = "{Swiftly Chasing Gravitational Waves across the Sky in Real Time}",
      journal = {\apjl},
         year = 2024,
        month = nov,
       volume = {975},
       number = {1},
          eid = {L19},
        pages = {L19},
          doi = {10.3847/2041-8213/ad87ce},
archivePrefix = {arXiv},
       eprint = {2410.05720},
 primaryClass = {astro-ph.HE},
       adsurl = {https://ui.adsabs.harvard.edu/abs/2024ApJ...975L..19T}
}

@ARTICLE{Barthelmy_2005_BAT,
       author = {{Barthelmy}, Scott D. and {Barbier}, Louis M. and {Cummings}, Jay R. and {Fenimore}, Ed E. and {Gehrels}, Neil and {Hullinger}, Derek and {Krimm}, Hans A. and {Markwardt}, Craig B. and {Palmer}, David M. and {Parsons}, Ann and {Sato}, Goro and {Suzuki}, Masaya and {Takahashi}, Tadayuki and {Tashiro}, Makota and {Tueller}, Jack},
        title = "{The Burst Alert Telescope (BAT) on the SWIFT Midex Mission}",
      journal = {\ssr},
         year = 2005,
        month = oct,
       volume = {120},
       number = {3-4},
        pages = {143-164},
          doi = {10.1007/s11214-005-5096-3},
archivePrefix = {arXiv},
       eprint = {astro-ph/0507410},
 primaryClass = {astro-ph},
       adsurl = {https://ui.adsabs.harvard.edu/abs/2005SSRv..120..143B}
}

@ARTICLE{Roming_2005_UVOT,
       author = {{Roming}, Peter W.~A. and {Kennedy}, Thomas E. and {Mason}, Keith O. and {Nousek}, John A. and {Ahr}, Lindy and {Bingham}, Richard E. and {Broos}, Patrick S. and {Carter}, Mary J. and {Hancock}, Barry K. and {Huckle}, Howard E. and {Hunsberger}, S.~D. and {Kawakami}, Hajime and {Killough}, Ronnie and {Koch}, T. Scott and {McLelland}, Michael K. and {Smith}, Kelly and {Smith}, Philip J. and {Soto}, Juan Carlos and {Boyd}, Patricia T. and {Breeveld}, Alice A. and {Holland}, Stephen T. and {Ivanushkina}, Mariya and {Pryzby}, Michael S. and {Still}, Martin D. and {Stock}, Joseph},
        title = "{The Swift Ultra-Violet/Optical Telescope}",
      journal = {\ssr},
         year = 2005,
        month = oct,
       volume = {120},
       number = {3-4},
        pages = {95-142},
          doi = {10.1007/s11214-005-5095-4},
archivePrefix = {arXiv},
       eprint = {astro-ph/0507413},
 primaryClass = {astro-ph},
       adsurl = {https://ui.adsabs.harvard.edu/abs/2005SSRv..120...95R}
}

@ARTICLE{Burrows_2005_XRT,
       author = {{Burrows}, David N. and {Hill}, J.~E. and {Nousek}, J.~A. and {Kennea}, J.~A. and {Wells}, A. and {Osborne}, J.~P. and {Abbey}, A.~F. and {Beardmore}, A. and {Mukerjee}, K. and {Short}, A.~D.~T. and {Chincarini}, G. and {Campana}, S. and {Citterio}, O. and {Moretti}, A. and {Pagani}, C. and {Tagliaferri}, G. and {Giommi}, P. and {Capalbi}, M. and {Tamburelli}, F. and {Angelini}, L. and {Cusumano}, G. and {Br{\"a}uninger}, H.~W. and {Burkert}, W. and {Hartner}, G.~D.},
        title = "{The Swift X-Ray Telescope}",
      journal = {\ssr},
         year = 2005,
        month = oct,
       volume = {120},
       number = {3-4},
        pages = {165-195},
          doi = {10.1007/s11214-005-5097-2},
archivePrefix = {arXiv},
       eprint = {astro-ph/0508071},
 primaryClass = {astro-ph},
       adsurl = {https://ui.adsabs.harvard.edu/abs/2005SSRv..120..165B}
}

@ARTICLE{BATPipeline_2025,
       author = {{Parsotan}, Tyler and {Palmer}, David M. and {Ronchini}, Samuele and {Delaunay}, James and {Tohuvavohu}, Aaron and {Laha}, Sibasish and {Lien}, Amy and {Cenko}, S. Bradley and {Krimm}, Hans and {Markwardt}, Craig},
        title = "{BatAnalysis{\textemdash}A Comprehensive Python Pipeline for Swift BAT Time-tagged Event Data Analysis}",
      journal = {\apj},
         year = 2025,
        month = jul,
       volume = {988},
       number = {1},
          eid = {23},
        pages = {23},
          doi = {10.3847/1538-4357/ade240},
archivePrefix = {arXiv},
       eprint = {2502.00278},
 primaryClass = {astro-ph.HE},
       adsurl = {https://ui.adsabs.harvard.edu/abs/2025ApJ...988...23P}
}

@ARTICLE{Evans_2023,
       author = {{Evans}, P.~A. and {Page}, K.~L. and {Beardmore}, A.~P. and {Eyles-Ferris}, R.~A.~J. and {Osborne}, J.~P. and {Campana}, S. and {Kennea}, J.~A. and {Cenko}, S.~B.},
        title = "{A real-time transient detector and the living Swift-XRT point source catalogue}",
      journal = {\mnras},
         year = 2023,
        month = jan,
       volume = {518},
       number = {1},
        pages = {174-184},
          doi = {10.1093/mnras/stac2937},
archivePrefix = {arXiv},
       eprint = {2208.14478},
 primaryClass = {astro-ph.HE},
       adsurl = {https://ui.adsabs.harvard.edu/abs/2023MNRAS.518..174E}
}

@ARTICLE{txs_2018,
       author = {{IceCube Collaboration} and {Aartsen}, M.~G. and {Ackermann}, M. and {Adams}, J. and {Aguilar}, J.~A. and {Ahlers}, M. and {Ahrens}, M. and {Al Samarai}, I. and {Altmann}, D. and {Andeen}, K. and {Anderson}, T. and {Ansseau}, I. and {Anton}, G. and {Arg{\"u}elles}, C. and {Auffenberg}, J. and {Axani}, S. and {Bagherpour}, H. and {Bai}, X. and {Barron}, J.~P. and {Barwick}, S.~W. and {Baum}, V. and {Bay}, R. and {Beatty}, J.~J. and {Becker Tjus}, J. and {Becker}, K.-H. and {BenZvi}, S. and {Berley}, D. and {Bernardini}, E. and {Besson}, D.~Z. and {Binder}, G. and {Bindig}, D. and {Blaufuss}, E. and {Blot}, S. and {Bohm}, C. and {B{\"o}rner}, M. and {Bos}, F. and {B{\"o}ser}, S. and {Botner}, O. and {Bourbeau}, E. and {Bourbeau}, J. and {Bradascio}, F. and {Braun}, J. and {Brenzke}, M. and {Bretz}, H.-P. and {Bron}, S. and {Brostean-Kaiser}, J. and {Burgman}, A. and {Busse}, R.~S. and {Carver}, T. and {Cheung}, E. and {Chirkin}, D. and {Christov}, A. and {Clark}, K. and {Classen}, L. and {Coenders}, S. and {Collin}, G.~H. and {Conrad}, J.~M. and {Coppin}, P. and {Correa}, P. and {Cowen}, D.~F. and {Cross}, R. and {Dave}, P. and {Day}, M. and {de Andr{\'e}}, J.~P.~A.~M. and {De Clercq}, C. and {DeLaunay}, J.~J. and {Dembinski}, H. and {De Ridder}, S. and {Desiati}, P. and {de Vries}, K.~D. and {de Wasseige}, G. and {de With}, M. and {DeYoung}, T. and {D{\'\i}az-V{\'e}lez}, J.~C. and {di Lorenzo}, V. and {Dujmovic}, H. and {Dumm}, J.~P. and {Dunkman}, M. and {Dvorak}, E. and {Eberhardt}, B. and {Ehrhardt}, T. and {Eichmann}, B. and {Eller}, P. and {Evenson}, P.~A. and {Fahey}, S. and {Fazely}, A.~R. and {Felde}, J. and {Filimonov}, K. and {Finley}, C. and {Flis}, S. and {Franckowiak}, A. and {Friedman}, E. and {Fritz}, A. and {Gaisser}, T.~K. and {Gallagher}, J. and {Gerhardt}, L. and {Ghorbani}, K. and {Glauch}, T. and {Gl{\"u}senkamp}, T. and {Goldschmidt}, A. and {Gonzalez}, J.~G. and {Grant}, D. and {Griffith}, Z. and {Haack}, C. and {Hallgren}, A. and {Halzen}, F. and {Hanson}, K. and {Hebecker}, D. and {Heereman}, D. and {Helbing}, K. and {Hellauer}, R. and {Hickford}, S. and {Hignight}, J. and {Hill}, G.~C. and {Hoffman}, K.~D. and {Hoffmann}, R. and {Hoinka}, T. and {Hokanson-Fasig}, B. and {Hoshina}, K. and {Huang}, F. and {Huber}, M. and {Hultqvist}, K. and {H{\"u}nnefeld}, M. and {Hussain}, R. and {In}, S. and {Iovine}, N. and {Ishihara}, A. and {Jacobi}, E. and {Japaridze}, G.~S. and {Jeong}, M. and {Jero}, K. and {Jones}, B.~J.~P. and {Kalaczynski}, P. and {Kang}, W. and {Kappes}, A. and {Kappesser}, D. and {Karg}, T. and {Karle}, A. and {Katz}, U. and {Kauer}, M. and {Keivani}, A. and {Kelley}, J.~L. and {Kheirandish}, A. and {Kim}, J. and {Kim}, M. and {Kintscher}, T. and {Kiryluk}, J. and {Kittler}, T. and {Klein}, S.~R. and {Koirala}, R. and {Kolanoski}, H. and {K{\"o}pke}, L. and {Kopper}, C. and {Kopper}, S. and {Koschinsky}, J.~P. and {Koskinen}, D.~J. and {Kowalski}, M. and {Krings}, K. and {Kroll}, M. and {Kr{\"u}ckl}, G. and {Kunwar}, S. and {Kurahashi}, N. and {Kuwabara}, T. and {Kyriacou}, A. and {Labare}, M. and {Lanfranchi}, J.~L. and {Larson}, M.~J. and {Lauber}, F. and {Leonard}, K. and {Lesiak-Bzdak}, M. and {Leuermann}, M. and {Liu}, Q.~R. and {Lozano Mariscal}, C.~J. and {Lu}, L. and {L{\"u}nemann}, J. and {Luszczak}, W. and {Madsen}, J. and {Maggi}, G. and {Mahn}, K.~B.~M. and {Mancina}, S. and {Maruyama}, R. and {Mase}, K. and {Maunu}, R. and {Meagher}, K. and {Medici}, M. and {Meier}, M. and {Menne}, T. and {Merino}, G. and {Meures}, T. and {Miarecki}, S. and {Micallef}, J. and {Moment{\'e}}, G. and {Montaruli}, T. and {Moore}, R.~W. and {Morse}, R. and {Moulai}, M. and {Nahnhauer}, R. and {Nakarmi}, P. and {Naumann}, U. and {Neer}, G.},
        title = "{Multimessenger observations of a flaring blazar coincident with high-energy neutrino IceCube-170922A}",
      journal = {Science},
         year = 2018,
        month = jul,
       volume = {361},
       number = {6398},
          eid = {eaat1378},
        pages = {eaat1378},
          doi = {10.1126/science.aat1378},
archivePrefix = {arXiv},
       eprint = {1807.08816},
 primaryClass = {astro-ph.HE},
       adsurl = {https://ui.adsabs.harvard.edu/abs/2018Sci...361.1378I}
}

@ARTICLE{Evans_2017,
       author = {{Evans}, P.~A. and {Cenko}, S.~B. and {Kennea}, J.~A. and {Emery}, S.~W.~K. and {Kuin}, N.~P.~M. and {Korobkin}, O. and {Wollaeger}, R.~T. and {Fryer}, C.~L. and {Madsen}, K.~K. and {Harrison}, F.~A. and {Xu}, Y. and {Nakar}, E. and {Hotokezaka}, K. and {Lien}, A. and {Campana}, S. and {Oates}, S.~R. and {Troja}, E. and {Breeveld}, A.~A. and {Marshall}, F.~E. and {Barthelmy}, S.~D. and {Beardmore}, A.~P. and {Burrows}, D.~N. and {Cusumano}, G. and {D'A{\`\i}}, A. and {D'Avanzo}, P. and {D'Elia}, V. and {de Pasquale}, M. and {Even}, W.~P. and {Fontes}, C.~J. and {Forster}, K. and {Garcia}, J. and {Giommi}, P. and {Grefenstette}, B. and {Gronwall}, C. and {Hartmann}, D.~H. and {Heida}, M. and {Hungerford}, A.~L. and {Kasliwal}, M.~M. and {Krimm}, H.~A. and {Levan}, A.~J. and {Malesani}, D. and {Melandri}, A. and {Miyasaka}, H. and {Nousek}, J.~A. and {O'Brien}, P.~T. and {Osborne}, J.~P. and {Pagani}, C. and {Page}, K.~L. and {Palmer}, D.~M. and {Perri}, M. and {Pike}, S. and {Racusin}, J.~L. and {Rosswog}, S. and {Siegel}, M.~H. and {Sakamoto}, T. and {Sbarufatti}, B. and {Tagliaferri}, G. and {Tanvir}, N.~R. and {Tohuvavohu}, A.},
        title = "{Swift and NuSTAR observations of GW170817: Detection of a blue kilonova}",
      journal = {Science},
         year = 2017,
        month = dec,
       volume = {358},
       number = {6370},
        pages = {1565-1570},
          doi = {10.1126/science.aap9580},
archivePrefix = {arXiv},
       eprint = {1710.05437},
 primaryClass = {astro-ph.HE},
       adsurl = {https://ui.adsabs.harvard.edu/abs/2017Sci...358.1565E}
}

@ARTICLE{Yi_2023,
       author = {{Yi}, Fang and {Haibo}, Yuan and {Ruoyi}, Zhang and {Jian}, Gao and {Shuai}, Xu},
        title = "{Empirical extinction coefficients for the Swift-UVOT optical-through-ultraviolet passbands}",
      journal = {\mnras},
         year = 2023,
        month = oct,
       volume = {525},
       number = {2},
        pages = {2701-2707},
          doi = {10.1093/mnras/stad2463},
archivePrefix = {arXiv},
       eprint = {2308.11664},
 primaryClass = {astro-ph.IM},
       adsurl = {https://ui.adsabs.harvard.edu/abs/2023MNRAS.525.2701Y}
}

@ARTICLE{Romano2006,
       author = {{Romano}, P. and {Campana}, S. and {Chincarini}, G. and {Cummings}, J. and {Cusumano}, G. and {Holland}, S.~T. and {Mangano}, V. and {Mineo}, T. and {Page}, K.~L. and {Pal'Shin}, V. and {Rol}, E. and {Sakamoto}, T. and {Zhang}, B. and {Aptekar}, R. and {Barbier}, S. and {Barthelmy}, S. and {Beardmore}, A.~P. and {Boyd}, P. and {Burrows}, D.~N. and {Capalbi}, M. and {Fenimore}, E.~E. and {Frederiks}, D. and {Gehrels}, N. and {Giommi}, P. and {Goad}, M.~R. and {Godet}, O. and {Golenetskii}, S. and {Guetta}, D. and {Kennea}, J.~A. and {La Parola}, V. and {Malesani}, D. and {Marshall}, F. and {Moretti}, A. and {Nousek}, J.~A. and {O'Brien}, P.~T. and {Osborne}, J.~P. and {Perri}, M. and {Tagliaferri}, G.},
        title = "{Panchromatic study of GRB 060124: from precursor to afterglow}",
      journal = {\aap},
         year = 2006,
        month = sep,
       volume = {456},
       number = {3},
        pages = {917-927},
          doi = {10.1051/0004-6361:20065071},
archivePrefix = {arXiv},
       eprint = {astro-ph/0602497},
 primaryClass = {astro-ph},
       adsurl = {https://ui.adsabs.harvard.edu/abs/2006A&A...456..917R}
}

@article{Roming_2009,
doi = {10.1088/0004-637X/690/1/163},
url = {https://doi.org/10.1088/0004-637X/690/1/163},
year = {2008},
month = {dec},
publisher = {The American Astronomical Society},
volume = {690},
number = {1},
pages = {163},
author = {Roming, P. W. A. and Koch, T. S. and Oates, S. R. and Porterfield, B. L. and Vanden Berk, D. E. and Boyd, P. T. and Holland, S. T. and Hoversten, E. A. and Immler, S. and Marshall, F. E. and Page, M. J. and Racusin, J. L. and Schneider, D. P. and Breeveld, A. A. and Brown, P. J. and Chester, M. M. and Cucchiara, A. and De Pasquale, M. and Gronwall, C. and Hunsberger, S. D. and Kuin, N. P. M. and Landsman, W. B. and Schady, P. and Still, M.},
title = {THE FIRST SWIFT ULTRAVIOLET/OPTICAL TELESCOPE GRB AFTERGLOW CATALOG},
journal = {The Astrophysical Journal},
}
\bibliographystyle{aasjournalv7}



\end{document}